\documentclass[prl,twocolumn,superscriptaddress,showpacs,preprintnumbers,amsmath,amssymb]{revtex4}
\usepackage{graphicx}% Include figure files
\usepackage{dcolumn}%Align table columns on decimal point
\usepackage{bm}% bold math
\usepackage{epsfig}
\newcommand{\ignore}[1]{}

\begin{document}

\title{Electronic Band Structure Mapping of Nanotube Transistors by Scanning Photocurrent Microscopy}

\author{Eduardo J. H. Lee}
\email{e.lee@fkf.mpg.de}
\address{Max-Planck-Institute for Solid State Research, Heisenbergstrasse 1, 70569 Stuttgart, Germany}
\author{Kannan Balasubramanian}%
\address{Max-Planck-Institute for Solid State Research, Heisenbergstrasse 1, 70569 Stuttgart, Germany}
\author{Jens Dorfm\"{u}ller}
\address{Max-Planck-Institute for Solid State Research, Heisenbergstrasse 1, 70569 Stuttgart, Germany}
\author{Ralf Vogelgesang}
\address{Max-Planck-Institute for Solid State Research, Heisenbergstrasse 1, 70569 Stuttgart, Germany}
\author{Nan Fu}
\address{University of Siegen, Adolf-Reichwein-Strasse 2, 57068 Siegen, Germany}
\author{Alf Mews}
\address{University of Siegen, Adolf-Reichwein-Strasse 2, 57068 Siegen, Germany}
\author{Marko Burghard}
\address{Max-Planck-Institute for Solid State Research, Heisenbergstrasse 1, 70569 Stuttgart, Germany}
\author{Klaus Kern}
\address{Max-Planck-Institute for Solid State Research, Heisenbergstrasse 1, 70569 Stuttgart, Germany}
\address{Institut de Physique des Nanostructures, \'{E}cole Polytechnique F\'{e}d\'{e}rale de Lausanne, CH-1015 Lausanne, Switzerland}

\date{\today}

\begin{abstract}
Spatially resolved photocurrent
measurements on carbon nanotube field-effect transistors
(CNFETs) operated in various transport regimes are reported. It is demonstrated
that the photocurrents measured at different biasing conditions
provide access to the electronic band structure
profile of the nanotube channel.  A comparison of the
profiles with the device switched into \emph{n}- or \emph{p}-type states
clearly evidences the impact of chemical doping from
the ambient. Moreover, we show that scanning photocurrent microscopy constitutes an effective
and facile technique for the quantitative determination of the Schottky
barrier height in such devices.
\end{abstract}


\maketitle

Carbon nanotube field-effect transistors (CNFETs) have been
extensively studied for applications in electronics and
optoelectronics \cite{cnfet, freitagpc}. Despite significant
progress in this research field, a range of important device
features remain to be fully elucidated, such as the nature of the
metal-CNT interface \cite{interface}. In this context, scanning
probe microscopies can provide relevant local information from
devices \cite{freitagsgm, defect1}. A well-suited technique to
explore the band structure profile within CNT devices is scanning
photocurrent microscopy (SPCM), which so far revealed dominant
photocurrent generation at the CNT-metal contacts \cite{bal1, bal2}.
Here we report the detailed SPCM characterization of CNFETs in
different charge transport regimes, and proof that optical
excitation of the nanotubes is the origin of the detected
photocurrents. The SPCM data enable convenient access to the height
of the Schottky barriers at the contacts. Furthermore, the obtained
band profiles are in excellent agreement with current models of
CNTFET device operation.

SWCNTs were synthesized via chemical vapor deposition (CVD),
following the procedure reported by Choi et al \cite{choi}. AuPd
electrical contacts (15nm thickness) were defined on top the tubes
using standard e-beam lithography. Photocurrent and reflection
images were simultaneously acquired through a confocal optical
microscope coupled to an electrical measurement set-up \cite{bal1}.
Photo-illumination was carried out by a HeNe laser (E $\sim$ 1.96
eV, spot size $\sim$ 0.5 $\mu$m) with a focused beam intensity of
$\sim$ 100 kW/cm$^2$. The photocurrent measurements were performed
under ambient conditions, on a total of 11 different devices.

Figure 1(a) shows the transfer characteristic of a CNFET displaying
slightly asymmetric ambipolar behavior. The photocurrent images of
the device taken at zero bias in both the \emph{p}- and
\emph{n}-type regimes, presented in Fig. 1(b), exhibit enhanced
photocurrent responses close to the metal contacts. By contrast, no
detectable photoresponse is observed along the entire CNFET device
in the OFF state. This difference can be explained by Schottky
barriers at the contacts prevalent in the ON states, whereas the
flat bands in the OFF-state do not support acceleration of the
photo-generated carriers [see insets of Fig. 1(b)]. The
gate-induced modulation of the SPCM images solidifies the previous
conclusion that built-in electric fields are responsible for the
photocurrent generation \cite{bal1, bal2}. Further support for such
interpretation stems from the sign of the photocurrent lobes in the
various regimes, as exemplified by the line profiles in Fig. 1(c).
Based upon the convention that electrons flowing out of the source
contact are measured as a positive current, the signs of the
photocurrents at the contacts in the ON states of the device are
consistent with the direction of the built-in fields there.

\begin{figure}
\includegraphics{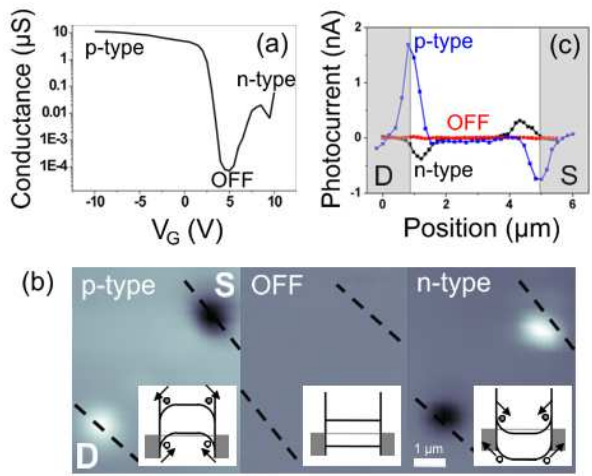}
\caption{\label{fig:epsart} (a) Conductance \emph{vs}. gate voltage
plot of a CNFET (b) Zero-bias SPCM images (white corresponds to
positive current). The dashed lines indicate the edges of the source
(S) and drain (D) electrical contacts. The insets depict the
respective band diagrams. (c) Photocurrent line profiles taken along
the nanotube.}
\end{figure}

Although the photoresponses in the \emph{p}- and \emph{n}-type
regimes are qualitatively similar, they exhibit some subtle
differences. While in the \emph{p}-type regime, the photocurrent
peaks appear almost exactly at the tube/contact interface, in the
\emph{n}-type regime they are offset by $\sim$ 0.25 $\mu$m towards
the center of the CNT channel. Such a behavior was observed in all
of the investigated samples. It demonstrates that the
band structure of the device in the \emph{p}-type regime is not just
a mirror image of that in the \emph{n}-type regime. To elucidate this further,
a series of zero-bias photocurrent images were acquired while sweeping the
gate voltage from the \emph{n}-type regime to the \emph{p}-type regime, as
depicted in Fig. 2(a). It is apparent that while tuning the device from the
\emph{n}-type regime to the OFF state, the photocurrent lobes shift
gradually towards the center of the channel, become broader and
considerably decrease in intensity. The very low intensity (few tens
of pA) features along the channel in the OFF state, observed for
some of the samples, may be ascribed to weak local electric fields
similar to those observed in metallic CNTs \cite{bal2}. Such fields
could arise due to, e.g. defect sites \cite{defect1}.
When the device is further tuned to the \emph{p}-type regime, the lobes
appear suddenly at the contacts without showing any gradual movements.
This observation is independent of the speed and direction of the gate voltage sweep.

The trend observed upon switching from the \emph{n}-type regime to
the OFF state is indicative of lowering and widening of the Schottky
barriers. These changes are consistent with theoretical simulations,
according to which the Schottky barrier width is closely related to
the magnitude of the local electric fields at the contacts
\cite{heinzeprl}. The sudden appearance of the \emph{p}-type lobes
in comparison to the gradual movement and broadening of the
\emph{n}-type lobes evidences that in the former regime the gate voltage
has a much weaker effect on the width of the Schottky barriers.
The overall behavior of the CNFET is in close accord with the
doped-CNFET model recently proposed by Chen and Fuhrer
\cite{chendoping}. It involves chemical \emph{p}-type doping upon
nanotube exposure to ambient conditions, thereby introducing a
strong band-bending confined to a very small width $W_{doped}$ for
\emph{p}-type operation [Fig. 2(b)] \cite{tersoffprl}. By
comparison, in the \emph{n}-type regime charge depletion occurs over
an effective barrier width $W_{eff}$ of the order of magnitude of
the gate oxide thickness \emph{t}, which can be effectively
modulated by the gate voltage.  This situation is different from
intrinsic-CNFETs for which in both regimes the bands are bent over a
depletion length \emph{W} $\sim$ \emph{t}. Within the doping model, the
occurrence of the \emph{p}-type photocurrent lobes at the contacts
and the $\sim$ 0.25 $\mu$m shift of the \emph{n}-type lobes [Figs.
1(c) and 2(a)] can be related to the presence of Schottky barriers with
respective widths of $W_{doped}$ and $W_{eff}$. Moreover, the
broadening of the lobes upon transition from the \emph{n}-type
regime to the OFF-state is attributable to an increase of $W_{eff}$,
whereas in the \emph{p}-type regime, the absence of such a behavior
indicates that the Schottky barrier width is confined to
$W_{doped}$. Therefore, it is evidenced that the effect of exposure
to ambient conditions is not restricted to changes in the metal work function.

Towards the task of determining the band structure profile of CNFETs
with the aid of SPCM it has to be ensured that the observed
photocurrent signals indeed originate from photoexcitation of the
nanotube. Previous CNFET photoconductivity studies have reported photovoltage
generation at the Si/SiO$_{2}$ interface upon excitation with energies above
the Si band gap \cite{freitagpc, marcusjap}. To rule out such
contributions in the SPCM response, measurements were performed in the \emph{p}-type
regime with laser excitation at $\lambda_{exc}$ = 1.6 $\mu$m (E $\sim$ 0.78 eV, spot size
$\sim$ 2 $\mu$m). The obtained SPCM images displayed the same principal
features as in the experiments with $\lambda_{exc}$ = 633 nm, proving
that the photoresponse predominantly arises from optical excitation
of the nanotubes.

\begin{figure}
\includegraphics{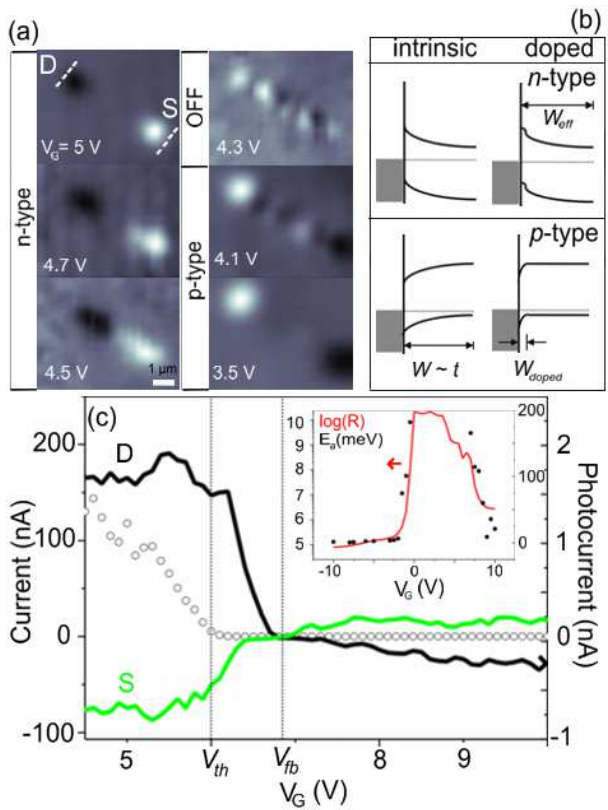}% Here is how to import EPS art
\caption{\label{fig:epsart}(a) Zero bias photocurrent response upon
transition from the \emph{n}- to \emph{p}-type. Color scales were
normalized such that white (black) corresponds to the maximum
(minimum) detected current. (b) Band diagrams depicting the
intrinsic- and doped-CNFET models. (c) Gate voltage dependence of
the drain current (dotted curve) and of the intensity of the
photocurrent lobes (solid lines). The inset shows the activation
barrier energy at zero bias (dots) and the logarithm of the
resistance (red line) as a function of gate voltage. $V_{th}$ and
$V_{fb}$ correspond to the threshold and flat-band voltages.}
\end{figure}

In order to determine the Schottky barrier height for holes $\Phi_{Bp}$,
the gate dependence of the drain current $I_{d}$ and the
photocurrent at the contacts $I_{ph}$ was evaluated, following a
similar procedure as reported for Si nanowires \cite{sinw}. The two
$I_{ph}$ curves in Fig. 2(c) cross each other at $V_{G}$ = $V_{fb}$
= 6.9 V, where $I_{ph}$ is approximately zero. At this gate voltage,
the device is in the OFF state with flat bands at the contacts. Moreover,
when $V_{G}$ equals the threshold voltage, the drain current reaches zero,
corresponding to the situation where the valence
band is approximately aligned at the Fermi level \cite{appenzeller}.
The Schottky barrier height can be calculated as $\Phi_{Bp} =
\alpha_{p}\cdot(V_{fb} - V_{th})$, where $\alpha_{p}$ is the gate
coupling parameter which can be estimated from the subthreshold
slope obtainable from the drain current curve in the \emph{p}-type
regime \cite{alpha}. For the sample in Fig. 2, the subthreshold
slope is $\sim$300 mV/decade, yielding an $\alpha_{p}$ value close to
0.2, from which $\Phi_{Bp}$ is determined to be $\sim$180 meV.
Similar values have been obtained for other samples.
In order to independently determine $\Phi_{Bp}$, we have performed
temperature dependent $I-V$ measurements and analyzed the data within
the framework of thermionic emission theory, as described in
previous works \cite{martelprl, chendoping}. The measured activation barrier energies at
zero bias are shown as a function of $V_{G}$ in the inset of Fig. 2(c).
In the ON states, the activation energy significantly underestimates the barrier
height, since charge injection occurs via thermal-assisted tunneling. On the
other hand, in the OFF state, thermionic emission is the dominant process, such
that the activation energy approaches $\Phi_{Bp}$. Due
to sensitivity limitations, a lower limit of $\sim$170 meV was
estimated for the barrier height, which is in close agreement to the value obtained by SPCM
measurements.

Having characterized the devices at zero bias, we now address the
effect of finite drain-source bias on the band structure profile of
CNFETs. Figure 3(a) displays a series of SPCM images taken at various
bias voltages with the device operated in the \emph{n}-type regime.
Although a somewhat smaller ratio of photocurrent to the dark
current is found in the \emph{p}-type regime, the following
discussion applies equally well in this case. The images disclose
that an applied voltage leads to the enhancement of one of the
photocurrent lobes, depending on the sign of the bias. For
sufficiently large bias, a single lobe is observed. To better visualize the CNFET
band profile, we integrate the photocurrent signal along the length
of the tube. This approach is justified by assuming that the
photocurrent is directly proportional to the built-in electric field
\cite{footnote}. In Fig. 3(b), the resulting qualitative
electrostatic potential profiles are depicted for different bias
values, with the source taken as ground and the drain being lifted
(lowered) upon application of a negative (positive) potential. It
follows from the line profiles that the center of the channels
remains approximately flat irrespective of the applied bias.
Moreover, application of a more positive (negative) bias results in
a voltage drop predominantly occurring at the source (drain). The
band profile changes with increasing bias closely following the
behavior expected for the \emph{n}-type unipolar operation mode of a CNFET,
confirming that the charge transport through the CNT is governed by
the contacts. Furthermore, the photocurrent at the lobes $I_{ph}$
is plotted as a function of $V_{ds}$ in Fig. 3(c). The $I_{ph}$-$V_{ds}$
curves are characteristic of reverse-biased Schottky diodes under
photoexcitation \cite{nearfield}.

\begin{figure}
\includegraphics{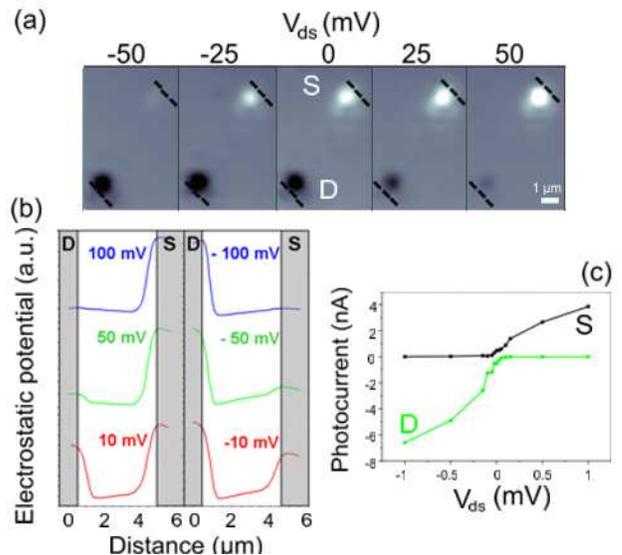}% Here is how to import EPS art
\caption{\label{fig:epsart}(a) Effect of bias voltage on the
photoresponse in the \emph{n}-type ON state. (b) CNFET electrostatic
potential profiles taken along the tube. (c) Bias dependence of the
photocurrent response at the contacts.}
\end{figure}

Finally, the effect of gate voltage on the photoresponse of biased
CNFETs was investigated. To this end, a series of images similar to
those in Fig. 2 were recorded at an applied bias of $V_{ds}$ =
+0.7 V, which are collected in Fig. 4(a). At the starting point of
$V_{G}$ = 10 V in the \emph{n}-type regime, a single lobe at the
source contact is observed, in close correspondence to the situation
in Fig. 3(a). Upon decreasing $V_{G}$, the intensity of this lobe
gradually decreases and clear photocurrent signals emerge along the
tube between the contacts. Around $V_{G}$ = 3.7 V,
\emph{p}-type operation sets in and a single lobe emerges at the drain
contact, analogous to the observation in Fig. 3(a). The
set of SPCM images provides a complete picture of the
photoconductivity in CNFETs, in close agreement with a
previously proposed model \cite{bal03}. A major conclusion is that
true photoconductivity from an individual CNT is only observable in
the OFF state of the device. In the ON states, the local photoresponse is
similar to that obtained from a reverse-biased Schottky diode.

\begin{figure}
\includegraphics{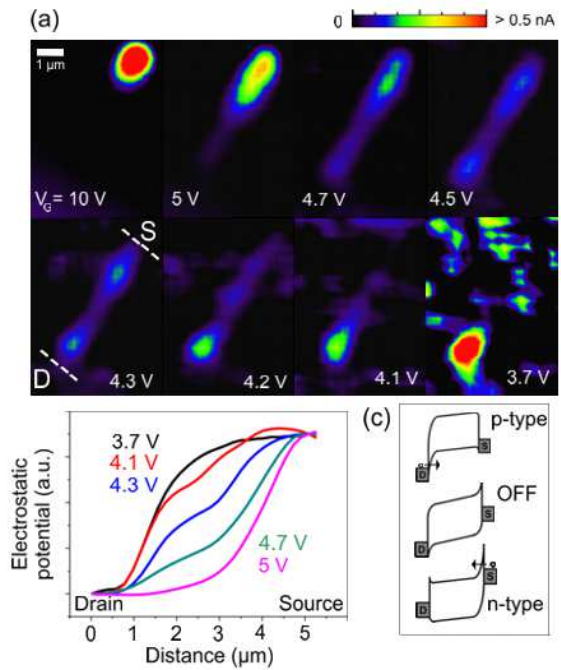}% Here is how to import EPS art
\caption{\label{fig:epsart}(a) SPCM images depicting the transition
from the \emph{n}- to the \emph{p}-type regimes taken at $V_{ds}$ =
0.7V. (b) Electrostatic potential profiles obtained from the
corresponding photocurrent images. (c) Band diagrams describing the
different transport regimes of a biased CNFET.}
\end{figure}

It can be recognized from the corresponding electrostatic potential
profiles in Fig. 4(b) that in the \emph{p}- (\emph{n}-) type regime,
the voltage drops are concentrated at the drain (source) contact, as
opposed to the OFF-state, in which the potential drop is distributed
along the entire channel. Closer inspection of the profile at
$V_{G}$ = 4.3 V suggests that the voltage drop occurring in the
vicinity of the contacts is symmetric and stronger than in the
middle of the CNT. In total, the band profiles agree very well with
the widely accepted model for the operating mechanism of CNFETs,
wherein the band structure at the contacts is modulated by the
fields induced by gate and drain-source voltages \cite{appenzeller}.
The band profiles are also consistent with the model used to interpret
electroluminescence experiments \cite{freitagprl}. Specifically, in
the OFF state, the voltage drops at the contacts are symmetric, thus
leading to balanced injection of electrons and holes, followed by
light emission from the center of the devices. Upon moving away from
the OFF state to the \emph{p}- (\emph{n}-) type regime, the
voltage is found to drop mostly at the drain (source) contact,
whereby hole (electron) injection is favored. Although the drain
bias used in the present experiment is not ideally suited for
observing electroluminescence (where $V_{ds}$ is required to be
twice as high as $V_{G}$), the extracted band profiles describe the
underlying scenario impressively well.

In summary, SPCM has been demonstrated to be a versatile tool for
CNFET characterization. The obtained photocurrent images confirm the
relevance of chemical doping in CNFETs. Moreover, it has been shown that
SPCM allows facile estimation of the Schottky barrier height, for whose
determination reliable and straightforward methods are still
lacking. The gained values of 150-200 meV are in good agreement with
theoretical predictions. Furthermore, SPCM experiments have revealed that
photoconductivity can be observed in the OFF state, whereas in the ON regimes
the local photoresponse is dominated by the contacts, which behave as
photoexcited reverse-biased Schottky diodes.

\bibliography{apssamp}% Produces the bibliography via BibTeX.

\begin{references}

\bibitem{cnfet}
Ph. Avouris, MRS Bull. \textbf{29}, 403 (2004).

\bibitem{freitagpc}
M. Freitag, Y. Martin, J. A. Misewich, R. Martel and Ph. Avouris,
Nano Lett. \textbf{3}, 1067 (2003).

\bibitem{interface}
Z. H. Chen, J. Appenzeller, J. Knoch, YM Lin, Ph. Avouris, Nano
Lett. \textbf{5}, 1497 (2005).

\bibitem{freitagsgm}
M. Freitag, M. Radosavljevic, Y. X. Zhou, A. T. Johnson, W. F.
Smith, Appl. Phys. Lett. \textbf{79}, 3326 (2001).

\bibitem{defect1}
M. Bockrath, W. Liang, D. Bozovic, J. H. Hafner, C. M. Lieber, M.
Tinkham and H. Park, Science \textbf{291}, 283 (2001).

\bibitem{bal1}
K. Balasubramanian, Y. Fan, M. Burghard, K. Kern, M. Friedrich, U. Wannek and A. Mews,
Appl. Phys. Lett. \textbf{84}, 2400 (2004).

\bibitem{bal2}
K. Balasubramanian, M. Burghard, K. Kern, M. Scolari, A. Mews,
Nano Lett. \textbf{5}, 507 (2005).

\bibitem{choi}
H. C. Choi, W. Kim, D. Wang and H. Dai,
J. Phys. Chem. B \textbf{106}, 12361 (2002).

\bibitem{heinzeprl}
S. Heinze, J. Tersoff, R. Martel, V. Derycke, J. Appenzeller and Ph. Avouris,
Phys. Rev. Lett. \textbf{89}, 106801 (2002).

\bibitem{chendoping}
Y. F. Chen and M. S. Fuhrer,
Nano Lett. \textbf{6} 2158 (2006).

\bibitem{tersoffprl}
F. Leonard and J. Tersoff,
Phys. Rev. Lett \textbf{89}, 179902 (2002).

\bibitem{marcusjap}
M. S. Marcus, J. M. Simmons, O. M. Castellini, R. J. Hamers and M. A. Eriksson,
J. Appl. Phys. \textbf{100}, 084306 (2006).

\bibitem{sinw}
Y. Ahn, J. Dunning and J. Park,
Nano Lett. \textbf{5}, 1367 (2005).

\bibitem{appenzeller}
J. Appenzeller, J. Knoch, V. Derycke, R. Martel, S. Wind and Ph. Avouris,
Phys. Rev. Lett. \textbf{89}, 126801 (2002).

\bibitem{alpha}
S. Rosenblatt, Y. Yaish, J. Park, J. Gore, V. Sazanova and P. L.
McEuen, Nano Lett. \textbf{2}, 89 (2002).

\bibitem{martelprl}
R. Martel, V. Derycke, C. Lavoie, J. Appenzeller, K. K. Chan, J. Tersoff and Ph. Avouris,
Phys. Rev. Lett. \textbf{87}, 256805 (2001).

\bibitem{footnote}
The electrostatic potential profile is valid only in the region between the contacts.
Moreover, it is broadened by convolution with the Airy pattern of the
incident laser beam.

\bibitem{nearfield}
Y. Gu, E. S. Kwak, J. L. Lensch, J. E. Allen, T. W. Odom and L. J. Lauhon,
Appl. Phys. Lett. \textbf{87}, 043111 (2005).

\bibitem{bal03}
K. Balasubramanian and M. Burghard,
Semicond. Sci. Technol. \textbf{21}, S22 (2006).

\bibitem{freitagprl}
M. Freitag, J. Chen, J. Tersoff, J. C. Tsang, Q. Fu, J. Liu and Ph. Avouris,
Phys. Rev. Lett. \textbf{93}, 076803 (2004).


\end{references}

\end{document}